# A Novel Structure of Advance Encryption Standard with 3-Dimensional Dynamic S-box and Key Generation Matrix

Ziaur Rahaman[1], Anjela Diana Corraya[2], Mousumi Akter Sumi[3], Ali Newaz Bahar[4]
Department of Information and Communication Technology
Mawlana Bhashani Science and Technology University
Tangail, Bangladesh
Email: zia@iut-dhaka.edu[1], anjela.it12012@gmail.com[2], sumiakterict@gmail.com[3], bahar_mitdu@yahoo.com[4]

*Abstract*— The study of sending and receiving secret messages is called cryptography. Generally, senders and receivers are unaware about the process of encryption and decryption. Hence, encryption plays an important role in data communication and data security. The meaning of encryption is not only to keep data confidential from unwanted access but also ensuring the data integrity through available way. As the capacity of breaking the security is increasing rapidly, so, the process that hides information is one of the most concerned topics. Advanced Encryption Standard is a popular, widely used and efficient encryption algorithm, which has been used since it was invented. This paper focuses on the AES key generation process and Substitution box. It modifies the conventional key generation technique and builds the dynamic 3-Dimensional S-box of Advance Encryption Standard. The proposed approach suggests 3-Dimensioanl Key Generation Matrix and S-box. As per shown this novel technique increases the amount of time it needs during encryption and decryption. The experimental result shows that it also enhances the strength of AES algorithm. The proposed approach illustrates the theoretical analysis and corresponding experimented results.

*Keywords—Advanced Encryption Standard; AES Modification; 3-dimensional Key Generation Matrix; dynamic S-box.*

## I. INTRODUCTION

The data transmission rate over the Internet has been getting massively increased. So in order to give full assurance over secured data transmission from sender to receiver is a great concern in this universe. Besides the confidentiality, data integrity is another important issue. Advanced Encryption Standard (AES) plays a vital rule to insure the data integrity and confidentiality. Rijndael is the original name of AES [1,2] which is established by National Institute of Standards and Technologies (NIST) [3]. Ciphers family includes different key and block sizes belongs to Rijndael [4]. AES [5] is a block cipher system divided into Diffusion and Confusion principles. In confusion, the length of plaintext and cipher text is same. But in diffusion the length of plaintext and cipher text is unequal. In enciphering system, key is the unavoidable part. So, in this paper, we have proposed a new modified key scheduling algorithm. On the other hand, AES is based on the S-box that increases the cryptographic strength. For this, we have generated dynamic 3-Dimensional S-box. In cryptanalysis, we know that Advanced Encryption Standard (AES) is generated from Galois Field GF ($2^8$) and introduced AES-128, AES-192 and AES-256. With observation it is clear that AES-192 is slower than AES-128 but AES-256 is more secure than AES 128. AES-256 is used to protect against quantum brute force attack.

The objective of this paper is to work with GF ($3^5$) which is feasible for 243 bits plaintext and 243 bits keys at a time. In order to modify traditional AES and to make it more efficient, proposed system has involved a 3-Dimensional Key Generation Matrix (3DKGM) for key generation and 3-Dimensional Dynamic S-box. The total number of round in this system is 16 which is divided into two parts named odd round (1,3,5,…,15) and even round (2,4,6,…,16). The main difference of these two rounds is absence of mix column in odd round and present in even round. So, it enhances complexity as well as complexity expensive for hackers. As a result, the proposed system stands as a secured system.

The rest of this paper is organized as follows: Issues and security in section 2, Related works in section 3, Describe problem statement in section 4, section 5 and 6 respectively go for 3-Dimensional Key Generation Matrix (3DKGM) System and Proposed 3-Dimensional Dynamic S-box, Section 7 represents the Proposed system, All experimental analysis and discussion take place in section 8 and finally Future perspective and conclusion in section 9.

## II. ISSUES IN SECURITY

The three security issues are: confidentiality, integrity and availability; known as the ACI triad [6].

*A. Availability*
The information that is formed and stored needs to be available to the authorized users. Without availability, the information is useless.

*B. Integrity*
Integrity assures that the information is changed by authorized entities and through authorized mechanisms. Unwanted change in information occurs risk to integrity. The sent data must be same as received data and not be altered through transmission path.

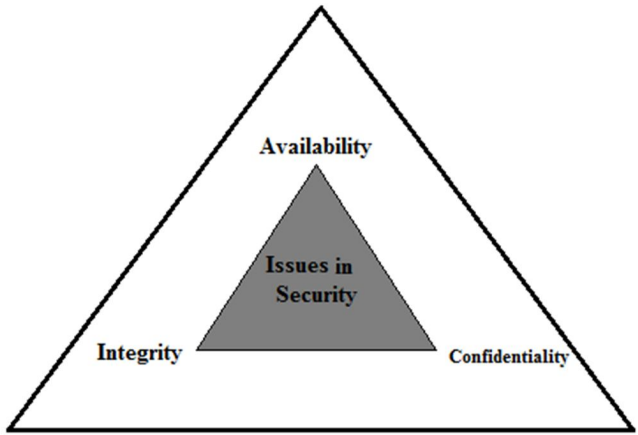

Fig. 1. Issues in Security.

*C. Confidentiality*

Data theft and unauthenticated access are raised [7] which can be protected by confidentiality. Confidentiality means to guard against danger that is ensured by data encryption services, authentication and security protocols.

### III. RELATED WORK

In this paper, we will just concentrate on the core components of AES key scheduling and AES S-box. Modification of these two properties is carried out in this paper. All the parts of modifications over AES is described here to make understand the differences of our proposed AES with traditional AES.

In 1997, a great deal of corporation is attack and in 1998 for implementation machines are too expensive. After that, with time computer's power is increasing and as a result there required stronger algorithm to face hackers attacks. Various works which are done earlier in AES field by many researchers. We analysis their work, idea, platform, limitations and also draw a survey in this section. In [5], they represent ASIC AES implementations of power analysis against the attack and extract information on side channel attack. But, at the time of simulating attack a huge number of noises are present. In [8], showed an image encryption algorithm for high definition image this is based on the modification of AES. But, it has an unstable round number at the time of attack and also the long time recruitments for the encryption and decryption process. In [9], they proposed an improved algebraic expression in the S-box generation which made the generation process more complicated. But, the limitation, excesses the computational cost over the improvement. In [6], several security issues are described that are concerned for cloud computing and also showed a number of serious security threats. In [10], a new system is proposed for data security called RSASS. RSA algorithm is used here to encrypting a large database or store data into any files. But limitation was this system is better for static data, but not better for linear methods with the retrieval speed.

For the experimental analysis of our key generation, a number of different file sizes are used to show their computational time and for the S-Box, total time is calculated for different number orientation. So with the analysis of result we can show that our approach is more efficient compared with other algorithm. The proposed system in this paper can provide a perfect combination of excellent security, efficiency, flexibility, implementability and performance.

### IV. PROBLEM STATEMENT

AES based on Rijndel Algorithm which is a standard combination of a strong algorithm and a strong key. With the calculation it can see that for AES-128 to check all possible key (50 billion keys per seconds) total required time is $5*10^{21}$ years[11]. as the steps of the proposed algorithm in this paper is more complex than traditional AES, so, the total required time is obviously more than the traditional one.

There are two options to ensure the security rate of any algorithm. These are based on time and cost. If an intruder's required time to break a system is greater than his life time, then that system is called a secure system.

And again if the total cost that required breaking the system is much more then it's initial making cost. Then the system is called secure over cost. Because for any intruder it's obviously not a choice to break the system which belongs much more cost then its initial rate.

With the observation of above drawbacks of traditional AES algorithm, this paper introduced a system to replace the 2-Dimensional process to 3-Dimensional process. This 3-Dimensional is used not only for key generation but also for S-box to make it dynamic. The step by step process of this system is capable for disarranging of initial message and key, that is enough to confuse the intruder. And this is happened by occurring bit operation complexity which leads to hackers drifting into undecidable problems.

### V. 3-DIMENSIONAL KEY GEMERATION MATRIX (3DKGM)

For the modified AES key generation, we used a 9×9×9 cube matrix which is shown at the Fig. 2. With this system, we can overcome the limitation to calculate $3^5$.

Our 3D matrix is a combination of Latin Alphabets (A-Z), integer value (0-9) and Greek symbols. So it can make the secrete massage more and more complex. The overview of this matrix can be seen on Fig. 2. To better understand, let consider a secret key: 'COMPLEX@+αµ' , which have to encrypt. At first of the process, the position of every byte is to be declared. In Table 1 the position for secrete key "COMPLEX@+αµ"is declare d.

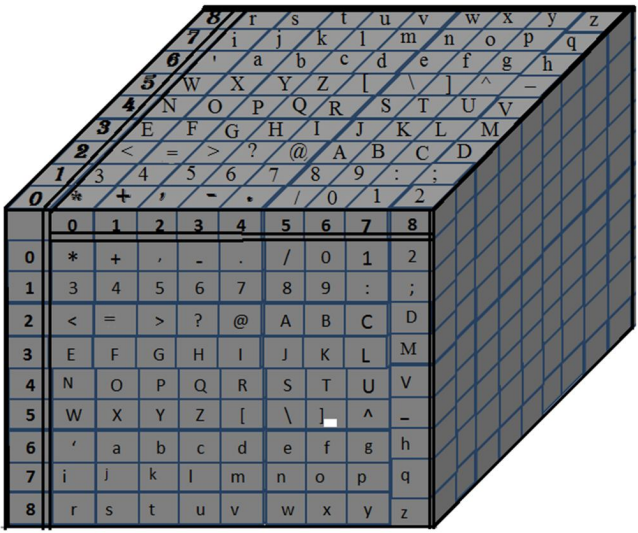

Fig. 2. Extended 3D Polybius 9×9×9 cube matrix.

### A. Encryption process in 3-Dimensional Key Generation Matrix System

To understand the encryption process of proposed 3DKGM system, we have shown an example in below. Here, it can be seen that from Fig. 3 for 'P', the row is first found (at x-axis) and the column (at y-axis) number from the 3D matrix.

TABLE I. POSITION OF SECRET KEY

| Secrete Message | C | O | M | P | L | X | @ | + | α |
|---|---|---|---|---|---|---|---|---|---|
| Position | 0 | 1 | 2 | 3 | 4 | 5 | 6 | 7 | 8 | 0 |

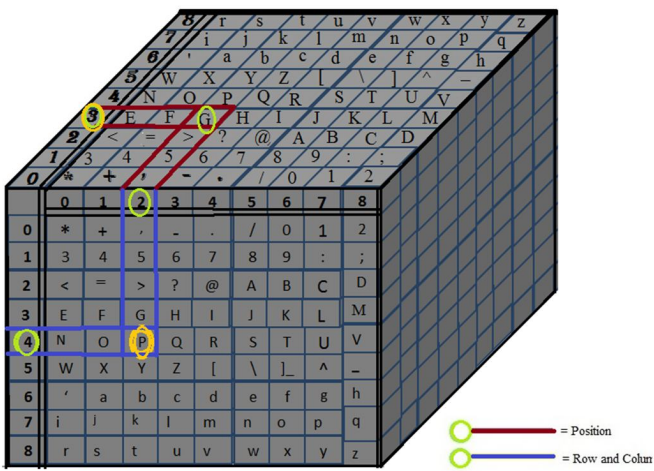

Fig. 3. Encryption procedure of 'P'

After that, at the z-axis we find the position of letter 'P'. So for "P" we can write "42G". So we get,

$$P = 42G \qquad (1)$$

Here, '4' for row, '2' for column and 'G' for position at 3 (can see from the Table 1).

With this encryption process the original text of key will be encrypted with some logical codes which is very much harder to identify. Here, the plain text is 'P' and the encrypted output for this '16G'.

### VI. PROPOSED 3-DIMENSIONAL DYNAMIC S-BOX

As discussed earlier, a dynamic 3-Dimensional S-box is generated. First, an initial S-box is needed which is 3-Dimensional and based on hexadecimal numbers. Fig. 4 and Fig. 6 are our generated initial 3-Dimensional S-box.

The 3-dimensional S-box is defined like: X (a, b, c) = Y1, Y2, Y3   Where,

a = index of x-axis

b = index of y-axis

c = index of z-axis

Y1 = Value of the row

Y2 = Value of the column

Y3 = Index value of c after selecting (Y1, Y2)

= (1st element of reversing hexadecimal number (next c follows the one byte cyclic left rotation of numbers), c, D) after selecting (Y1, Y2)

D = Hexadecimal number starts from (16/2+1)th value and run up to last and again in a cyclic order starts from the first to (16/2)th value (follows the number cycling).

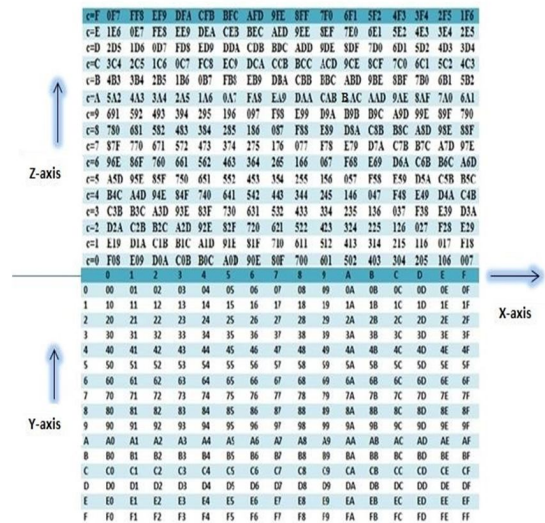

Fig. 4. 3-Dimensional S-box

For better understanding, let us take an example: the XOR-ed result of plaintext and key is 'ABC'. Now, from the above Fig. 4, For 'A' we go through the x-axis and then y-axis for 'B'. Finally, for 'C' we go through z-axis and select the value corresponding x, y and z-axis which is 'BCC'. In traditional S-box, hardware implementation is too hard. But, in proposed S-box, we can avoid these complexities.

*A. DynamicSbox Algorithm*

A word called dynamic is also used for 3-dimensional S-box. So, after getting the total initial keys, a random number will be generated. According to the random number, the keys will be 'left-rotated'. The value of rotation will be stored in a variable and according to this value; the S-box will be rotated again. Fig. 5 is showing an algorithm of dynamic S-box. For

```
void DynamicSbox(s_box,key){
    Random random = newRandom();
    intrandomInt = random.nextInt(16);
    int Count = GetShiftCount(randomInt);
    for(int i = count; i <= 16; i++){
        s_box=s_box[y3]+i;}}
```

Fig. 5. An algotithm of dynamic S-box

example, if the Count =2 (rotation is 2 times for keys), then the value of dynamic 3-Dimensioanl S-box is given in Fig. 7.

The random numbers and the rotation of S-box depend on the number 16 as the total number of hexadecimal is 16. If the shifting value in each z-axis is greater than 16, then no rotation is occurred in initial 3-Dimensional S-box. The rotation will occur in z-axis only as it is the main part of dynamic 3-Dimensional S-box.

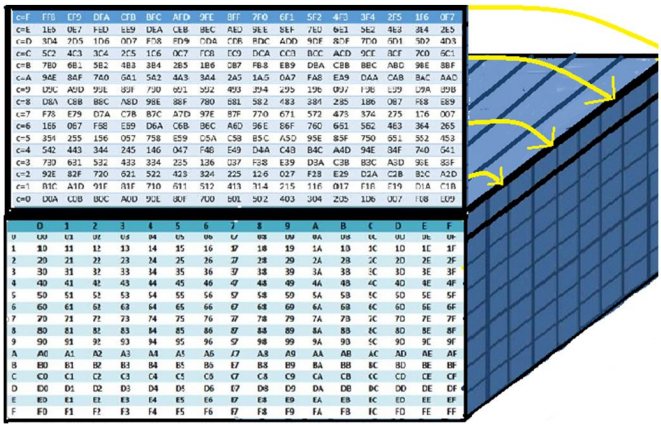

Fig. 6. 3-Dimensional S-box.

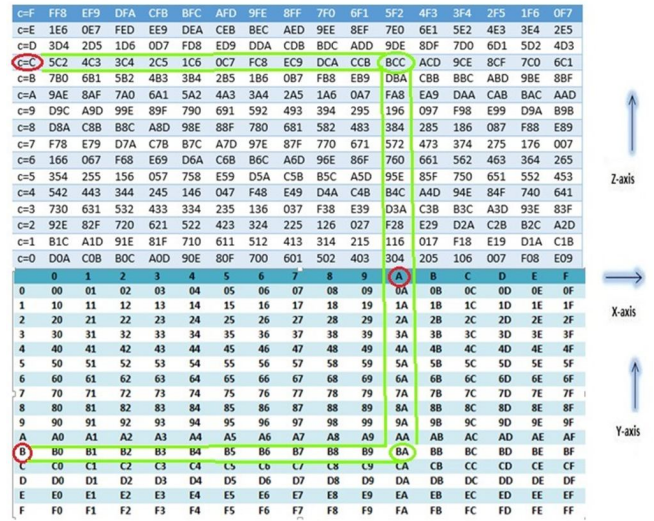

Fig. 7. 3-DimensionalS-box when rotation value is 2.

VII. PROPOSED SYSTEM

The Science of information transmiting and retrieving securelyover the insecure channel is called Cryptography [12]. There are two parts in the study of cryptography. They are: encryption and decryption. The 1st part is Encryption where a sender is converting data into an unintelligible string or cipher text during transmission, so that a hacker could not know about the sent data. And the 2nd part ,Decryption is just the reverse of it. With a proper decryption process the receiver converts sender's cipher text into a plaintext [13], as a meaningful text.

In the AES algorithm, the 1st step is the XOR operation of Plaintext and the key of same length of bits [3]. So in this work, we have focused on the key generation method and next is an S - box. We know, the background of AES depends on Galois Field. The working procedure of the system is given as follow;

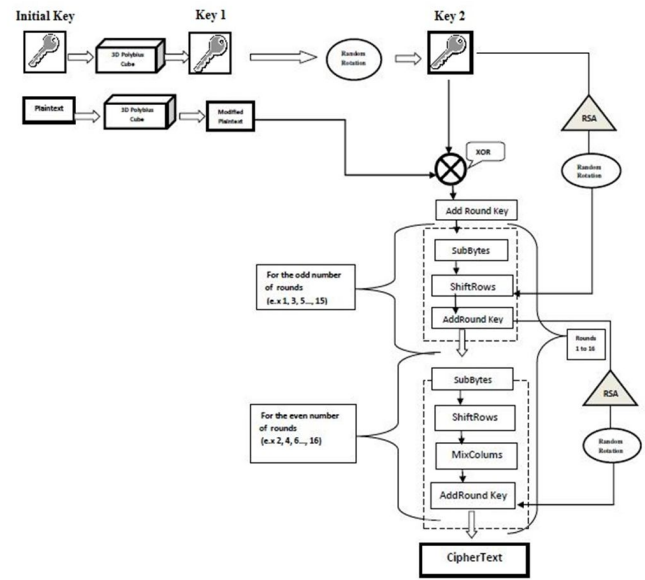

Fig. 8. The Modified AES Encryption Algorithm.

- To follow the rules of Galois Field, we have to take $(P)^n$ number of bits as a key and as a plaintext. Where, P has to be a prime number and n to be any integer value. So we used here 243 bits ($3^5$). But to convert these 243 bits into bytes (hexadecimal), we perform zero padding (5 zeros) at the last place of the key. Total 248 bits are obtained here (243+5=248); now we convert the bits into byte and gets 31 bytes (248÷8=31).

- In the next step, the proposed 3DKGM is used and 3 byte is obtained for every 1 byte. Details are described about our proposed 3-Dymensional Key Generation matrix system in the previous section. So, after using the logical calculations of 3-Dimensional matrix, we get 93 bytes (31×3=93) means 744 bits.

- Now, to make these 744 bits more complex and also for matching with the proposed S-box, used random rotation method. In random rotation the bit will be rotated towards left at anti-clockwise.

- For plaintext, the same procedures used for transforming 243 ($3^5$) bits into 744 bits. After this step, we have XOR-ed these (744 bits) key with the Plaintext (744 bits). And the XOR resulted will be forwarded for next steps where S-box exists.

- The resulted bits will be forwarded to the round procedures. That is already mentioned above sections (16 rounds). These 16 rounds are partitioned into odd round and even round.

- Even round procedure includes the SubBytes, ShiftRows, MixCloums, AddRound Key.

- Odd round procedure includes the SubBytes, ShiftRows, AddRound Key.

The main different between the two round procedure is absent of Mixcloumns in the odd round.

It is already known that key generation process never create any bad impacts over the hardware implementation. In order to increase strength of key, the key generation process was tried to make harder and also S-box exists in SubBytes. In the Fig. 8, we can see the overall schema of our proposed modified AES algorithm.

*A. The Output -Code Snippet for 3-Dynamices Key Generation Matrix (3DKGM)Algorithm*

Here in Fig. 9, Fig. 10 and Fig. 11, we can see some parts of the output for our key generation process using Java. In Fig. 9 we see the output for the message "Information and Communication Technology" using 3DKGM algorithm, which is our plaintext in the system.

In Fig. 10, we can see some output matrix parts of the 3DKGM algorithm. As the length of the total matrix is so long, so we skip the middle parts in this paper for proper representation.

```
Finding the plainText using 3DKGM
33E70+70,70-
70.50/68x61:82D82<81481583H77m23J23K33L61;84N
65X65Y87l76d83J83K64U76h75W65X23G23H
The time for generating plainText using 3DKGM
is  17 millisecond
```

Fig. 9. Output for the PlainText using 3DKGM

In Fig.11, here is shown the output for finding key using the 3DKGM algorithm.

After all these process, our next task is the rotation method. These step by step processes are shown before in the flow diagram on Fig. 8.

Form the Fig 8, we can see that next we perform the XOR-operation between the key and the modified Plaintext in our system. Then we are gone for further tasks, those belongs to the 3-Dymentional S-box.

```
The 3D Key Generating Matrix...
arr[0][0][0] = **         arr[0][0][1] = *+
arr[0][1][0] = +3         arr[0][1][1] = +4
arr[0][2][0] = ,<         arr[0][2][1] = ,=
arr[0][3][0] = -E         arr[0][3][1] = -F
arr[0][4][0] = .N         arr[0][4][1] = .O
arr[0][5][0] = /W         arr[0][5][1] = /X
arr[0][6][0] = 0`         arr[0][6][1] = 0a
arr[0][7][0] = 1i         arr[0][7][1] = 1j
arr[0][8][0] = 2r         arr[0][8][1] = 2s
         .
         .
         .
arr[8][0][7] = r1         arr[8][0][8] = r2
arr[8][1][7] = s:         arr[8][1][8] = s;
arr[8][2][7] = tC         arr[8][2][8] = tD
arr[8][3][7] = uL         arr[8][3][8] = uM
arr[8][4][7] = vU         arr[8][4][8] = vV
arr[8][5][7] = w^         arr[8][5][8] = w_
arr[8][6][7] = xg         arr[8][6][8] = xh
arr[8][7][7] = yp         arr[8][7][8] = yq
arr[8][8][7] = zy         arr[8][8][8] = zz
```

Fig. 10. Output for the 3D key generating Matrix.

```
Finding the key using 3DKGM
34N75X66b76c80.74S61982C70276`75X25Y75Z64R27n
76f74U74V83E75X70,63H61782A70076g75_46`65X63G
68u
The time for generating key using 3DKGM is
26 millisecond
```

Fig. 11. Output for key using 3DKGM.

## VIII. EXPERIMENTAL ANALYSES AND DISCUSSION

Time to encrypt and decrypt is an important feature of any encryption algorithm. As, Key scheduling and S-box are the parts of encryption algorithm, so, time has great impact on these.

As because of dynamic 3-dimensional S-box, the strength of our proposed algorithm is increased. In security analysis, it will take too long time for brute force approach. Wadi and Zainal recently proposed a S-box based on modified AES-128 [3] block cipher which is too easy to break . We did two types of experiment and simulated on Matlab2010a and we solve our algorithm with Java ()

*A. Computational Time vs. File Size*

In this part we showed a comparison result of our proposed approach with others 3 algorithms (AES, DES and TDES). We used different size of file and calculate the computational time. We know that the computational time for encryption process is the total time for the algorithm to convert plaintext into cipher text. In we can calculate the performance by calculation different time take by different algorithm.

TABLE II. COMPUTATIONAL TIME FOR ENCRYPTION VS. FILE SIZE IN KB

| File Sixe in KB | Computational Time for Encryption in Sec. | | | |
|---|---|---|---|---|
| | *AES* | *DES* | *TDES* | *Our Proposed* |
| 20 | 26 | 25 | 27 | 28 |
| 35 | 53 | 56 | 64 | 58 |
| 155 | 251 | 270 | 287 | 261 |
| 100 | 436 | 448 | 476 | 468 |
| 300 | 436 | 448 | 476 | 468 |
| 512 | 469 | 481 | 509 | 501 |

Here from the table II, we can see that, we take different file size of 20, 35, 155, 333 and 512.

And for 20kb we get the execution time 26, 25, 27, and 28 for the AES, DES, TDES and our proposed algorithm. Similarly, for 333Kb file size it gives 469,481, 509 and 501 for the AES, DES, TDES and our proposed algorithm.

So from the following table, we can see that, our proposed approach decrease the computational time rather than others algorithm. As a result it cans build up more security for the vast system.

From the table II, we draw a graph in order to show the vast dependency of computational time over file size. So from the above table, we showed the graphical representation in Fig. 12. Here it can be seen that the proposed approach has lower computational time compared to other algorithms for sophisticated bits. So finally we can say that our proposed algorithm is efficient than others algorithm from the comparison results of Table II and graph.

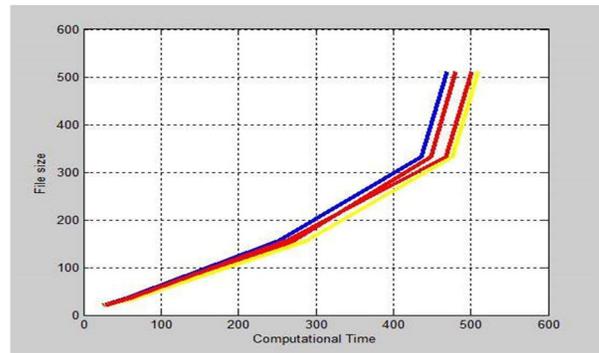

Fig. 12. Computational Time for Encryption in Sec. .

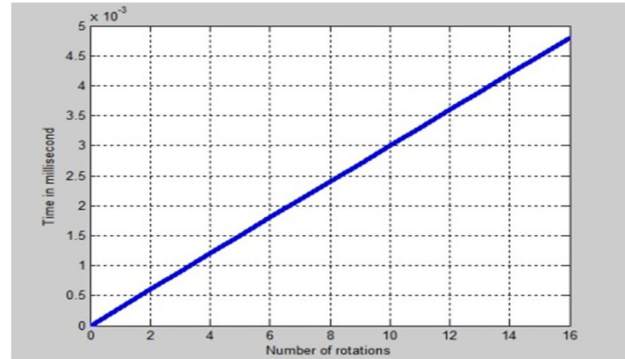

Fig. 13. Graphical representation of number of rotation vs. time (m sec.).

*B. Time in numbers of rotations*

Wadi and Zainal in [14] recently proposed an S-box based on modified AES-128 block cipher which is too easy to break which is claimed in [14]. So, the proposed 3-Dimensional S-box is dynamic, it depends on rotation. Rotation also takes some times, which has a great impact on computational time. Table III is showing the time in millisecond for rotations where rotation is possible from 0 to 16.In security analysis, it will take too long time for brute force approach.

TABLE III. TIME IN MILLISECOND FOR NUMBERS OF ROTATIONS IN Y AXIS

| Number of Rotation | Times(msec.) |
|---|---|
| 0 | 0.000 |
| 1 | 0.003 |
| 2 | 0.006 |
| 3 | 0.009 |
| 4 | 0.012 |
| 5 | 0.015 |
| 6 | 0.018 |
| 7 | 0.021 |
| 8 | 0.024 |
| 9 | 0.027 |
| 10 | 0.030 |
| 11 | 0.033 |
| 12 | 0.036 |
| 13 | 0.039 |
| 14 | 0.042 |
| 15 | 0.045 |
| 16 | 0.048 |

From the above table, it is observed that time is reasonable for any kind of rotation over S-box as it is dynamic. From the table III, we draw a graph on Fig. 13 in order to show the graphical representation of number of rotation vs. time in milliseconds.

### C. Average time for different Galois Field (GF)

Time to encrypt and decrypt is an important feature of any encryption algorithm. As, S-box is a part of encryption algorithm, so, time has great impact on this dynamic 3-dimesional S-box. As we work in GF($3^5$), we want to show the time of other n=1,2,3,4,5 which is less than 6 because $3^6$ creates some additional complexity which cannot be solved. Table IV is showing the average time of different GF($3^n$) where n is an integer and is less than 6.

TABLE IV. TIME IN MILLISECOND OF GF($3^n$) WHEN N<6

| Length of bits | Average time to generate dynamic 3-dimensional S-box (ms) |
|---|---|
| 3 | 0.0003 |
| 9 | 0.0057 |
| 81 | 0.0285 |
| 243 | 0.057 |

From the table IV, we draw a graph on Fig. 14 in order to show the graphical representation of time vs. length of bits.

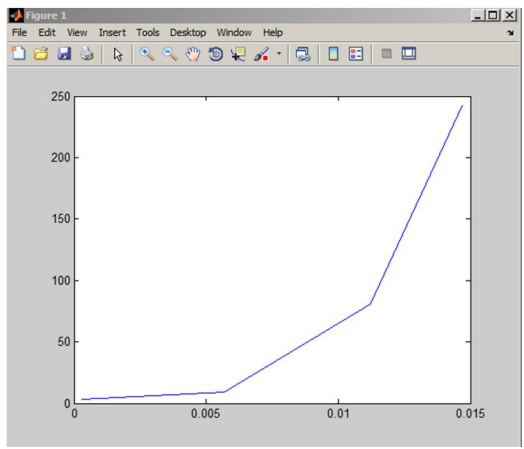

Fig. 14. Graphical representation of time (ms) vs. length of bits

### IX. FUTURE PERSPECTIVES AND CONCLUSION

Many attempts [15] have been made to modify AES algorithm based on GF($2^8$). All of these modifications did not show the computational time of rounds and intermediate tasks on which the feasibility of AES depends. As it is known that key scheduling does not effect on time of encryption and decryption, we can make complex calculations. But in this paper, we mainly focus both on key scheduling as well as S-box. A dynamic 3-dimensional S-box is created based on multiplicative polynomial inverse over GF ($3^5$) with 243 bits plaintext. In which, a new Substitution box (S-box) is proposed. For better security, we made it little complex keep pace with the computational time which is not so high. By thinking about security, we believe these dynamic 3-dimensional S-box and 3-dimensional cube key generation process can be used instead of traditional S-box.

In future, this system will be devolved to the image encryption standard based with 3-dymensional process. Up Next task of this system would be adding the authentication part for data security over cloud computing. At that stage the system will be concerned about the performance.